\documentclass[english,preprint,showpacs,preprintnumbers,amsmath,amssymb]{revtex4}
\usepackage[T1]{fontenc}
\usepackage[latin9]{inputenc}
\usepackage{amsmath}
\usepackage{amssymb}
\usepackage{epsfig}

\makeatletter
\@ifundefined{textcolor}{}
{%
 \definecolor{BLACK}{gray}{0}
 \definecolor{WHITE}{gray}{1}
 \definecolor{RED}{rgb}{1,0,0}
 \definecolor{GREEN}{rgb}{0,1,0}
 \definecolor{BLUE}{rgb}{0,0,1}
 \definecolor{CYAN}{cmyk}{1,0,0,0}
 \definecolor{MAGENTA}{cmyk}{0,1,0,0}
 \definecolor{YELLOW}{cmyk}{0,0,1,0}
 }

\RequirePackage{color}

\def\be{\begin{equation}}
\def\ee{\end{equation}}

\def\bea{\begin{eqnarray}}
\def\eea{\end{eqnarray}}

\makeatother

\usepackage{amssymb}
\usepackage{amsmath}
\usepackage{graphicx}

\usepackage{babel}

\begin{document}

\title{The dimer model on the triangular lattice.}

\author{N. Sh. Izmailian}
\email{izmailan@phys.sinica.edu.tw} \affiliation{Institute of
Physics, Academia Sinica, Nankang, Taipei 11529, Taiwan}
\affiliation{Alikhanyan National Science Laboratory, Alikhanian
Brothers 2, 375036 Yerevan, Armenia}

\author{Ralph Kenna}
\email{r.kenna@coventry.ac.uk} \affiliation{Applied Mathematics
Research Centre, Coventry University, Coventry CV1 5FB, England}

\date{\today}

\begin{abstract}
We analyze the partition function of the dimer model on an
$\mathcal{M} \times \mathcal{N}$ triangular lattice wrapped on
torus obtained by Fendley, Moessner and Sondhi [Phys. Rev. B
\textbf{66}, 214513 (2002)]. From a finite-size analysis we have
found that the dimer model on such a lattice can be described by
conformal field theory having central charge $c=-2$. The shift
exponent for the specific heat is found to depend on the parity of
the number of lattice sites $\mathcal{N}$ along a given lattice
axis: e.g., for odd $\mathcal{N}$ we obtain the shift exponent
$\lambda=1$, while for even $\mathcal{N}$ it is infinite
($\lambda=\infty$). In the former case, therefore, the finite-size
specific-heat pseudocritical point is size dependent, while in the
latter case, it coincides with the critical point of the
thermodynamic limit.
\end{abstract}


\maketitle

%%%%%%%%%%%%%%%%%%%%%%%%%%%%%%%%%%%%%%%%%%%%%%%%%%%%%%%%%%%%%%%%%%%%%%%%%%%%%%%%%%%%%%%%%%%%%%%%%%%%%%%%%%%%%%%%%%%%%%%%%%%%%%%%%%%%%
\section{Introduction}
\label{introduction}
%%%%%%%%%%%%%%%%%%%%%%%%%%%%%%%%%%%%%%%%%%%%%%%%%%%%%%%%%%%%%%%%%%%%%%%%%%%%%%%%%%%%%%%%%%%%%%%%%%%%%%%%%%%%%%%%%%%%%%%%%%%%%%%%%%%%%

In experiments and in numerical studies of critical phenomena, it
is essential to take into account finite-size effects in order to
extract correct infinite-volume predictions from the data.
Therefore, in recent decades there have been many investigations
of finite-size scaling, finite-size corrections, and boundary
effects for critical model systems. In the quest to improve our
understanding of realistic systems of finite extent,
two-dimensional models play a crucial role in statistical
mechanics as they have long served as a testing ground to explore
the general ideas of finite-size scaling under controlled
conditions. Very few of them have been solved exactly, the Ising
model
\cite{ferdinand1969,JaKe02,izmailian2001,izmailian2002b,izmailian2002,izmailian2002a,Kenna2001,Kenna2002,Kenna2002a}
and the dimer model
\cite{ferdinand1967,izmailian2003,izmailian2005,kong2006,kong2006a,izmailian2006}
being the most prominent examples.

The dimer model is a two-particle system. The main difference
between it and one-particle systems  such as Ising, Heisenberg, or
Potts models etc., is that occupation of a given lattice site
ensures that at least one of its nearest-neighbor sites is also
occupied. It is well known that, due to this non-locality, the
critical behavior exhibited by the dimer model can depend on the
lattice structure and shape (square, triangle, honeycomb, etc).
Previous studies have shown that finite-size corrections in the
free energy can exhibit a strong dependence upon the parity of the
lattice, and this has provoked controversial conclusions about the
value of the central charge from $c = -2$ to $c = 1$. One can
expect that such unusual finite-size behavior should also hold for
the dimer model on other lattices and here we investigate the
triangular lattice in particular.

We are particularly interested in the finite-size scaling behavior
of the specific-heat pseudocritical point. In finite systems the
counterparts of the singularities which mark higher-order phase
transitions in the thermodynamic limit are smooth peaks the shapes
of which depend on the critical exponents. In particular, let
$C(t,L)$ be the specific heat at a reduced temperature given by
$t$ for a system of linear extent characterized by $L$. In the
infinite-volume limit, $C(t,\infty)$ diverges at the critical
point $t=t_c=0$. In finite volume, the analog to the divergence is
a finite peak the shape of which is characterized by (i) its
position $t_{\rm{pseudo}}$ (ii) its height $C(t_{\rm{pseudo}},L)$
and (iii) its value at the infinite-volume critical point
$C(0,L)$. In particular, the position of the specific heat peak,
$t_{\rm{pseudo}}$, is a pseudocritical point which approaches
$t_c=0$ as $L^{-\lambda}$, where $\lambda$ is called the shift
exponent. In most models exhibiting higher-order phase
transitions, the shift exponent coincides with the inverse of the
correlation-length critical exponent $1/\nu$, but this is not a
direct conclusion of FSS theory is not always true.

For example, for the Ising model in two dimensions, Ferdinand and
Fisher determined that behavior of the specific-heat
pseudocritical point matches that of the correlation length with
$\lambda = 1/\nu = 1$ \cite{ferdinand1969}. However, Ising models
defined on two-dimensional lattices with other topologies have
shift exponents which differ from the inverse correlation length
critical exponent (see Ref.\cite{JaKe02} and references therein).
This is despite the fact that the critical properties on such
lattices are the same as for the torus in the thermodynamic limit.
A question we wish to address here is the corresponding status of
the shift exponent in the dimer model.

In contrast to spin models, the critical behavior of dimer models
are strongly influenced by the structure of the underlying
lattice. For example the square lattice dimer model is critical
with algebraic decay of correlators \cite{Fisher2,Fisher3}, while
the dimer model on the anisotropic honeycomb lattice, which is
equivalent to five-vertex model on the square lattice \cite{Wu5},
exhibits a KDP-type singularity and the dimer model on the
Fisher-type lattice exhibits Ising-type transitions
\cite{Fisher1}. Thus, it appears that the dimer model itself has
not a single critical behavior, but several critical behaviors
associated with different classes of universality.

It has been shown explicitly \cite{Kasteleyn} that the free energy
per site for the dimer model on the square lattice is insensitive
to the precise form of the boundary conditions in the limit of a
large lattice. This is in contrast to its finite-size counterpart,
for which sensitivity to boundary conditions is notable feature,
in particular to the parity of the number of lattice sites along a
given lattice axis \cite{izmailian2003,Ferdinand}. Similar
statements  hold for the dimer model on the honeycomb and
triangular lattices.

Very recently, it has been shown \cite{izmailian2005} that the
finite-size corrections of the dimer model on planar $\infty
\times \mathcal{N} $ square lattices also depend crucially on the
parity of $\mathcal{N} $ and the boundary conditions and such
unusual finite-size behavior can be fully explained in the
framework of the $c=-2$ logarithmic conformal field theory.

Our objective in this paper is to study the finite size properties
of dimer model on the plane triangular lattice using the same
techniques developed in papers
\cite{izmailian2003,izmailian2002b}. The paper is organized as
follows. In Sec. II we introduce the dimer model on the triangular
lattice with periodic boundary conditions. In Sec. III we discuss
the finite-size corrections for an infinitely long cylinder of
circumference $\mathcal{N}$ and find that the dimer model on the
triangular lattice can be described by conformal field theory with
central charge $c=-2$. In Sec. IV we investigate the properties of
the specific heat near the critical point and find that the
specific-heat  shift exponent $\lambda$ depends on the parity of
the number of lattice sites along the lattice axis $\mathcal{N}$.
For odd $\mathcal{N}$ we obtain for the shift exponent
$\lambda=1$, while for even $\mathcal{N}$ we find that the shift
exponent $\lambda$ is infinity ($\lambda=\infty$).
 Our results are summarized and discussed in Sec. V.

%%%%%%%%%%%%%%%%%%%%%%%%%%%%%%%%%%%%%%%%%%%%%%%%%%%%%%%%%%%%%%%%%%%%%%%%%%%%%%%%%%%%%%%%%%%%%%%%%%%%%%%%%%%%%%%%%%%%%%%%%%%%%%%%%%%%%
\section{Partition Function}
%%%%%%%%%%%%%%%%%%%%%%%%%%%%%%%%%%%%%%%%%%%%%%%%%%%%%%%%%%%%%%%%%%%%%%%%%%%%%%%%%%%%%%%%%%%%%%%%%%%%%%%%%%%%%%%%%%%%%%%%%%%%%%%%%%%%%

In the present work, we consider the dimer model on $\mathcal{M}
\times \mathcal{N}$ triangular lattice under periodic boundary
conditions. The partition function is given by
\begin{figure}[tbp]
\includegraphics[width=0.42\textwidth]{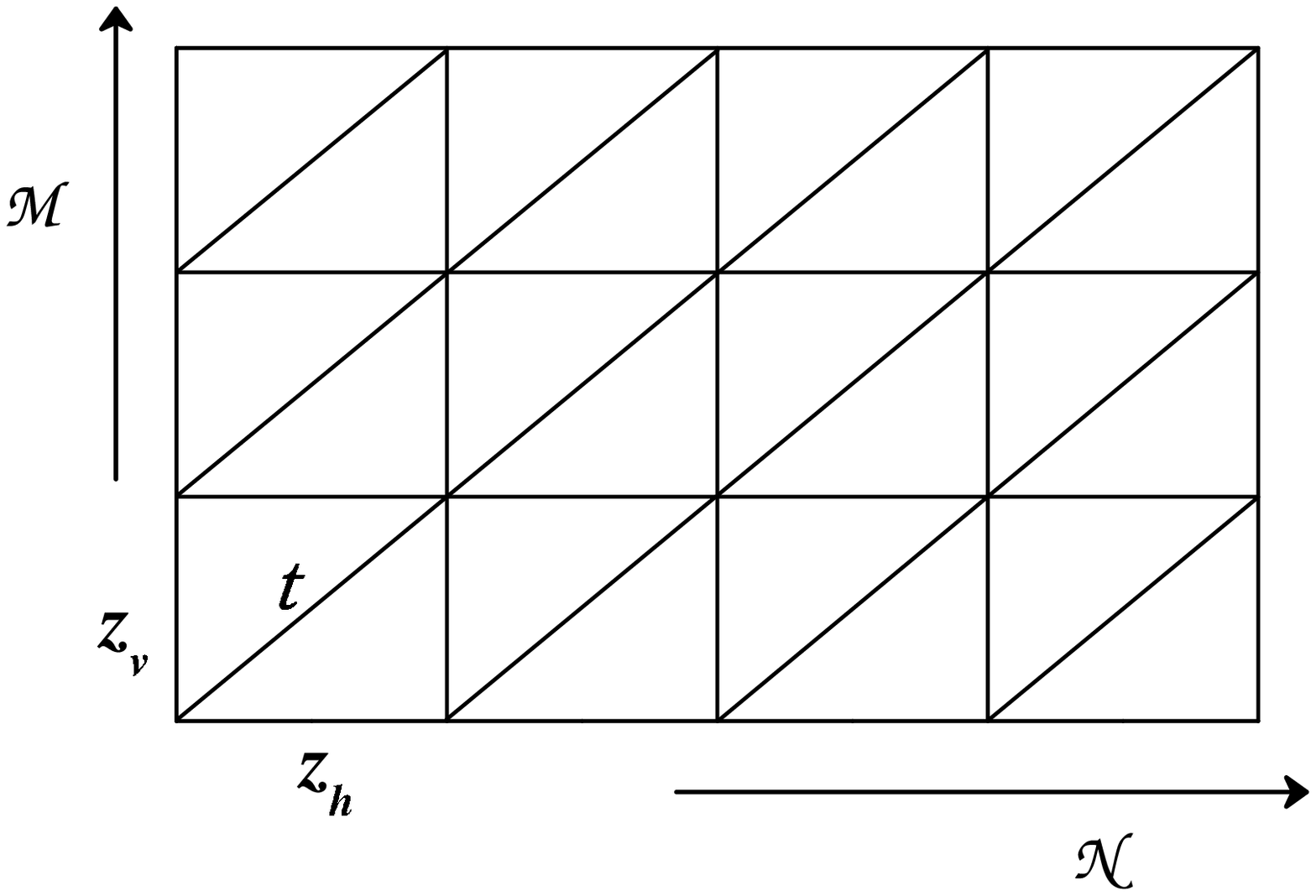}
\caption{The $\mathcal{M} \times \mathcal{N}$ triangular lattice
with dimer weights $z_h$ in the horizontal direction, $z_v$ in the
vertical direction, and $t$ in the diagonal direction.}
\label{fig1}
\end{figure}
\begin{equation}
Z_{{\mathcal{M}},{\mathcal{N}}}(z_h, z_v,t)=\sum
z_h^{n_h}z_v^{n_v}t^{n_t}, \label{PartitionFunctionDimer}
\end{equation}
where summation is taken over all dimer covering configurations,
where $z_h$, $z_v$ and $t$ are, respectively, dimer weights in the
horizontal, vertical and diagonal directions, and where $n_h$,
$n_v$ and $n_t$ are, respectively, the number of horizontal,
vertical and diagonal dimers (Fig.\ref{fig1}). The dimer model on
the triangular lattice undergoes phase transition at the point
$t=t_c=0$ (and likewise for $z_h$ and $z_v$), where the partition
function is singular. Thus the general triangular lattice model is
critical in the square lattice limit. The dimer weight $t$ plays a
role similar to the reduced temperature in the Ising model. In
what follows, we will set $z_h = z_v = 1$.

An explicit expression for the partition function of the dimer
model on an $\mathcal{M} \times \mathcal{N}$ triangular lattice
wrapped on torus has been obtained by Fendley, Moessner and Sondhi
\cite{Moesner} and can be written as \cite{izmailian2006}
\begin{equation}
Z_{\mathcal{M},
\mathcal{N}}(t)=\frac{1}{2}\left[G_{0,0}(t,\mathcal{M},
\mathcal{N})+G_{0,1/2}(t,\mathcal{M},
\mathcal{N})+G_{1/2,0}(t,\mathcal{M},
\mathcal{N})+G_{1/2,1/2}(t,\mathcal{M}, \mathcal{N})\right],
\label{partf}
\end{equation}
where
\begin{eqnarray}
G_{\alpha ,\beta }^{2}(t,\mathcal{M}, \mathcal{N}) & = &
\prod_{m=0}^{\mathcal{M}/2-1} \prod_{n=0}^{\mathcal{N}-1}4 \left[{
\sin ^{2}\frac{2\pi (n+\alpha )}{\mathcal{N}} }\right.
\nonumber \\
& & \left.{+\sin ^{2}\frac{2\pi (m+\beta )}{\mathcal{M}}+t^{2}\cos
^{2}\left( \frac{2\pi (n+\alpha )}{\mathcal{N}}+\frac{2\pi
(m+\beta )}{\mathcal{M}} \right) }\right] , \label{dt}
\end{eqnarray}
for even $\mathcal{M}$. The notation $\alpha =0$ corresponds to
periodic boundary conditions for the underlying free fermion in
the $\mathcal{N}$-direction while $\alpha =\frac{1}{2}$ represents
anti-periodic boundary conditions. The boundary conditions in the
$\mathcal{M}$-direction are similarly controlled by the parameter
$\beta $.

Since the total number of sites must be even if the lattice is to
be completely covered by dimers, we will consider two cases,
namely even-even (ee) case when $\mathcal{M}=2 M$ and
$\mathcal{N}=2 N$, and even-odd (eo) case when $\mathcal{M}=2 M$
and $\mathcal{N}=2 N+1$. Note that due to the symmetry of the
lattice the odd-even case (oe) ($\mathcal{M}=2 M +1$,
$\mathcal{N}=2N$) can be obtained from even-odd case by simple
transformation $\xi \rightarrow 1/\xi$, where
$\xi=\mathcal{M}/\mathcal{N}$ is a aspect ratio.

\vspace{0.5 cm}

\textbf{a). Dimers on $2 M \times 2 N$ lattices}
%%%%%%%%%%%%%%%%%%%%%%%%%%%%%%%%%%%%%%%%%%%%%%%%%%%%%%%%%%%%%%%%%%%%%%%%%%%%%%%%%%5

\vspace{0.3 cm}

In the even-even case where $(\mathcal{M},\mathcal{N}) = (2
M,2N)$, the second product in (\ref{dt}) may be compactly written
as $\displaystyle{\prod_{n=0}^{2N-1}F(n+\alpha, m+\beta)}$ where
the function $F(x,y)$ is given by
\begin{equation}
F(x,y)=4\left[\sin ^{2}\frac{\pi x}{N} +\sin ^{2}\frac{\pi y
}{M}+t^{2}\cos ^{2}\left( \frac{\pi x}{N}+\frac{\pi y}{M} \right)
\right]. \label{Fx}
\end{equation}
Splitting this product into two parts,
\begin{equation}
\prod_{n=0}^{2N-1} F(n+\alpha,y)=\prod_{n=0}^{N-1} F(n+\alpha,y)
\times \prod_{l=N}^{2N-1} F(l+\alpha,y), \label{Fx1}
\end{equation}
shifting the index in the second part from $l$ to $n=l-N$, and
noting the translation symmetry $F(N+x)=F(x)$, this may be
expressed as
\begin{equation}
\prod_{n=0}^{2N-1}F(n+\alpha,y)=
\left(\prod_{n=0}^{N-1}F(n+\alpha,y)\right)^2. \label{kap}
\end{equation}
Defining the partition function  with twisted boundary conditions
\begin{eqnarray}
 Z_{\alpha ,\beta}^{2}(t,M,N) & = & \prod_{m=0}^{M-1}\prod_{n=0}^{N
-1}4\left[{\sin ^{2}\frac{\pi (n+\alpha )}{N}}\right.
\nonumber \\
& & \left.{+\sin ^{2}\frac{\pi (m+\beta )}{M}+t^{2}\cos ^{2}\left(
\frac{\pi (n+\alpha )}{N}+\frac{\pi (m+\beta )}{M} \right)
}\right] ,
 \label{ab}
\end{eqnarray}
one has
\begin{equation}
G_{\alpha ,\beta }(t,2M,2N)=Z_{\alpha ,\beta}^{2}(t,M,N).
\end{equation}
The even-even partition function given by Eq.(\ref{partf}) and
Eq.(\ref{dt}) can now be written in the  form \cite{izmailian2006}
\begin{equation}
Z_{2M,2N}(t)=\frac{1}{2}\left[
Z_{0,0}^{2}(t,M,N)+Z_{0,1/2}^{2}(t,M,N)+Z_{1/2,0}^{2}(t,M,N)+Z_{1/2,1/2}^{2}(t,M,N)
\right] .  \label{ee}
\end{equation}
Note that Eq.(\ref{ab}) at $t=0$ coincides with the corresponding
expressions for the square lattice for which a general theory
about its asymptotic expansion has been given in
Ref.\cite{izmailian2002b}.

Note also that for all $(\alpha, \beta) \ne (0,0)$, the partition
function $Z_{\alpha ,\beta }(t,M,N)$ is even with respect to its
argument t. Hence, near the critical point ($t=0$) we have
\begin{eqnarray}
Z_{\alpha ,\beta }(t,M,N)&=&Z_{\alpha ,\beta
}(0,M,N)+\frac{t^2}{2}Z{''}_{\alpha ,\beta }(0,M,N)+... \qquad
\mbox{for} \quad (\alpha, \beta) \ne (0,0). \label{ident1}
\end{eqnarray}
The only exception is the point where both $\alpha$ and $\beta$
are equal to zero. This case has to be treated separately since at
this point ($t=0$) the partition function $Z_{0,0}(t,M,N)$
vanishes. As a result, we have
\begin{eqnarray}
Z_{0,0}(t,M,N)=t Z^{'}_{0,0
}(0,M,N)+\frac{t^3}{6}Z{'''}_{0,0}(0,M,N)+...\qquad \mbox{for}
\quad (\alpha, \beta) = (0,0). \label{ident2}
\end{eqnarray}
At the critical point $t=0$ we have
\begin{eqnarray}
Z_{0,0}(0,M,N)&=& 0\label{A100}\\
Z_{\alpha ,\beta }(0,M,N)&=&\prod_{n=0}^{N-1}\left\vert 2\sinh
\left[ M \omega\left(\frac{\pi (n+\alpha)}{N}\right)+i\pi \beta
\right] \right\vert \qquad \mbox{for} \quad (\alpha, \beta) \ne
(0,0), \label{A1}
\end{eqnarray}
where $\omega (k)=\mathrm{arcsinh}\left( \sin {k}\right)$. In the
derivation of the Eq.(\ref{A1}) we have used the identity
\cite{GradshteinRyzhik}
\begin{equation}
\prod_{m=0}^{M-1}4\textstyle{ \left[~\!{\sinh}^2\omega +
\sin^2\left(\frac{\pi
(m+\beta)}{M}\right)\right]}=4\left|~\!{\sinh}\left(M\omega+i\pi\beta\right)\right|^2.
\label{identity}
\end{equation}

 Taking the derivative of Eq.(\ref{ab}) with respect to
variable $t$ and then considering limit $t \to 0$ we obtain
\begin{eqnarray}
Z_{0,0}^{\prime }(0,M,N)&=&2M\prod_{n=1}^{N-1}2\left\vert \sinh
(M\omega{\left({\frac{\pi n}{N}}\right)} )\right\vert, \\
Z_{\alpha ,\beta }^{\prime }(0,M,N)&=& 0 \qquad \mbox{for} \quad
(\alpha, \beta) \ne (0,0).
\end{eqnarray}

\vspace{0.5 cm}

\textbf{b). Dimers on $2 M \times (2 N + 1)$ lattices}
%%%%%%%%%%%%%%%%%%%%%%%%%%%%%%%%%%%%%%%%%%%%%%%%%%%%%%%%%%%%%%%%%%%%%%%%%%%%%%%%%%5

\vspace{0.3 cm}

In Ref.\cite{izmailian2006}, it has been shown that in the
even-odd case, the partition function given by Eq.(\ref{partf})
and Eq.(\ref{dt}) can be written as
\begin{equation}
Z_{2M,2N+1}(t) = Z_{0,0}(t,M,2N+1)+Z_{0,1/2}(t,M,2N+1). \label{eo}
\end{equation}

Note that Eqs. (\ref{ee}) and (\ref{eo}) in the case $t=0$
coincide with the corresponding expressions for the square lattice
(see Ref.\cite{izmailian2003}).

Thus we can see that partition function for the dimer model on
triangular lattice under periodic boundary conditions can be
expressed in terms of the only one subject, namely,
$Z_{\alpha,\beta}(t,M,N)$ with $(\alpha, \beta) = (0,0), (0,1/2),
(1/2,0)$ and $(1/2,1/2)$.

%%%%%%%%%%%%%%%%%%%%%%%%%%%%%%%%%%%%%%%%%%%%%%%%%%%%%%%%%%%%%%%%%%%%%%%%%%%%%%%%%%%%%%%%%%%%%%%%%%%%%%%%%%%%%%%%%%%%%%%%%%%%%%%%%%%%%
\section{Dimer on the infinitely long strip}
\label{inf_long}
%%%%%%%%%%%%%%%%%%%%%%%%%%%%%%%%%%%%%%%%%%%%%%%%%%%%%%%%%%%%%%%%%%%%%%%%%%%%%%%%%%%%%%%%%%%%%%%%%%%%%%%%%%%%%%%%%%%%%%%%%%%%%%%%%%%%%

Conformal invariance of the model in the continuum scaling limit
would dictate that the asymptotic finite-size scaling behavior of
the critical free energy  of an infinitely long two-dimensional
strip of finite width ${\mathcal{N}}$ has the form
\begin{equation}
f=f_{\rm{bulk}}+\frac{2f_{\rm{surf}}}{\mathcal{N}}  +
\frac{A}{{\mathcal{N}}^2} + ..., \label{freeenergystrip}
\end{equation}
where $f_{\rm{bulk}}$ is the bulk free energy, $f_{\rm{surf}}$ is a
surface free energy and $A$ is constant. Unlike the free energy
densities $f_{bulk}$ and $f_{\rm{surf}}$, the constant A is universal.
The value of $A$ is  related to the central charge $c$ and the
highest conformal weight $\Delta$ of the underlying conformal
theory, and depends on the boundary conditions in the transversal
direction. These two dependencies combine into a function of the
effective central charge  $c_{eff} = c-24 \Delta$
\cite{Blote,Affleck,Cardy},
\begin{eqnarray}
A&=&-\frac{\pi}{24}c_{eff}= \pi\left(\Delta-\frac{c}{24}\right)
\qquad \mbox{on a strip,} \label{Astrip}\\
A&=&-\frac{\pi}{6}c_{eff}= 4\pi\left(\Delta-\frac{c}{24}\right)
\qquad \mbox{on a cylinder.} \label{Aperiod}
\end{eqnarray}
Let us now consider the dimer model on the infinitely long strip
of width $\mathcal{N}$ under periodic boundary conditions.

Considering the logarithm of the partition function given by
Eq.(\ref{A1}), we note that it can be transformed as
\begin{equation}
\ln Z_{\alpha,\beta}(0,M,N)= M\sum_{n=0}^{N-1}
\omega\!\left(\textstyle{\frac{\pi(n+\alpha)}{N}}\right)+
\sum_{n=0}^{N-1}\ln\left|\,1-e^{-2\big[\,M
\omega\left(\frac{\pi(n+\alpha)}{N}\right)-i\pi\beta\,\big]}\right|.
\label{lnZab}
\end{equation}
The second sum here vanishes in the formal limit $M\to\infty$. The
asymptotic expansion of the first sum can be found with the help
of the Euler-Maclaurin summation formula
\begin{equation}
M\sum_{n=0}^{N-1}\omega\!\left(\textstyle{\frac{\pi(n+\alpha)}{N}}\right)=
\frac{S}{\pi}\int_{0}^{\pi}\!\!\omega(x)~\!{\rm
d}x-\pi\lambda_0\rho\,{\rm B}_{2}^\alpha-
2\pi\rho\sum_{p=1}^{\infty} \left(\frac{\pi^2\rho}{S}\right)^{p}
\frac{\lambda_{2p}}{(2p)!}\;\frac{{\rm B}_{2p+2}^\alpha}{2p+2},
\label{EulerMaclaurinTerm}
\end{equation}
where $\int_{0}^{\pi}\!\!\omega(x)~\!{\rm d}x = 2 \gamma$, $\gamma
=0.915965...$ is Catalan's constant and ${\rm B}^{\alpha}_{p}$ are
so-called Bernoulli polynomials and $S=M N$. We have also used the
symmetry property, $\omega(k)=\omega(\pi-k)$, of the lattice
dispersion relation $\omega(k)$ and its Taylor expansion
\begin{equation}
\omega(k)=\sum_{p=0}^{\infty}
\frac{\lambda_{2p}}{(2p)!}\;k^{2p+1},
\label{SpectralFunctionExpansion}
\end{equation}
where $\lambda_0=1$, $\lambda_2=-2/3$, $\lambda_4=4$, etc.

Thus one can easily write down all the terms of the exact
asymptotic expansion for the $F_{\alpha,\beta}(N)=-\lim_{M \to
\infty}\frac{1}{M}\ln Z_{\alpha,\beta}(M,N)$
\begin{eqnarray}
F_{\alpha,\beta}(N) &=&-\lim_{M \to \infty}\frac{1}{M}\ln
Z_{\alpha,\beta}(M,N)= -\frac{2\gamma}{\pi}N +2\sum_{p=0}^\infty
\left(\frac{\pi}{N}\right)^{2p+1}\frac{\lambda_{2p}}{(2p)!}
\frac{B_{2p+2}^{\alpha}}{2p+2}. \label{AsymptoticExpansion1}
\end{eqnarray}

From $F_{\alpha,\beta}(N)$, we can obtain the asymptotic expansion
of the free energy per bond of an infinitely long cylinder of
circumference $\mathcal{N}$.  Since the expression for the
partition function is different for even $\mathcal{N}$ and odd
$\mathcal{N}$, we will consider these two cases separately. For
even $\mathcal{N}$ ($\mathcal{N}=2N$), we have
\begin{equation}
f = -\lim_{M \to \infty} \frac{1}{4 M N}\ln{Z_{2M,2N}(0)} =
-\lim_{M \to \infty} \frac{1}{2 M N}\ln {Z_{1/2, 0}(M,
N)}=\frac{1}{2 N} F_{1/2,0}(N) , \label{free2N}
\end{equation}
and for odd $\mathcal{N}$
\begin{eqnarray}
f &=& -\lim_{M \to \infty} \frac{1}{2M(2N+1)}\ln{Z_{2M, 2N+1}(0)}
= -\lim_{M \to \infty}\frac{1}{2M(2N+1)}\ln {Z_{0,
1/2}(M,2N+1)}\nonumber\\
&=&\frac{1}{2(2N+1)}F_{0,1/2}(2N+1). \label{free2N1}
\end{eqnarray}

From Eq.(\ref{free2N}) using Eq.(\ref{AsymptoticExpansion1}) one
can easily obtain that for even ${\mathcal{N}}$ the asymptotic
expansion of the free energy is given by
\begin{eqnarray}
f&=&f_{\rm{bulk}} +\frac{1}{\pi}\sum_{p=0}^\infty \left(\frac{2
\pi}{\mathcal{N}}\right)^{2p+2}\frac{\lambda_{2p}}{(2p)!}
\frac{B_{2p+2}^{1/2}}{2p+2}
\nonumber\\
&=& f_{\rm{bulk}}-\frac{\pi}{6}\frac{1}{\mathcal{N}^2}+\dots \quad
({\rm for}~ \mathcal{N}=2N), \label{2Nper}
\end{eqnarray}
while for odd ${\mathcal{N}}$ from Eqs.(\ref{free2N1}) and
(\ref{AsymptoticExpansion1}) one can obtain
\begin{eqnarray}
f&=&f_{\rm{bulk}} +\frac{1}{\pi}\sum_{p=0}^\infty
\left(\frac{\pi}{\mathcal{N}}\right)^{2p+2}\frac{\lambda_{2p}}{(2p)!}
\frac{B_{2p+2}}{2p+2} \nonumber\\
&=& f_{\rm{bulk}}+\frac{\pi}{12}\frac{1}{\mathcal{N}^2}-\dots
\quad ({\rm for} ~\mathcal{N}=2N+1). \label{2N1per}
\end{eqnarray}
The bulk free energy $f_{\rm{bulk}} = -\frac{\gamma}{\pi}$ is the
same for $\mathcal{N}$ even and odd cases. Thus we find that the
finite-size corrections in a crucial way depend on the parity of
$\mathcal{N}$. In particular it means that due to the certain
non-local features present in the dimer model, a change of parity
of $\mathcal{N}$ induces a change of boundary condition. The
similar situation also happen in the dimer model on the square
lattice, see \cite{izmailian2005}, where a detailed analysis of
boundary conditions and parity dependence effects has been carried
out in this context.

%In particular, it has been shown that by changing variables from
%dimer coverings to spanning trees that a change of parity of
%$\mathcal{N}$ has precisely the effect of changing the boundary
%conditions.

Since the effective central charge merely determines some
combination of $c$ and $\Delta$, one cannot obtain the values of
both without some assumption about one of them. This assumption
can be a posteriori justified if the conformal description
obtained from it is fully consistent. Surprisingly, there are two
consistent values of $c$ that can be used to describe the dimer
model, namely $c=-2$ and $c=1$. For example for the dimer model on
an infinitely long cylinder of even circumference $\mathcal{N}$
one can obtained from Eqs.(\ref{freeenergystrip}), (\ref{Aperiod})
and (\ref{2Nper}) that the central charge $c$ and the highest
conformal weight $\Delta$ can take values $c = 1$ and $\Delta=0$
or $c=-2$ and $\Delta=-1/8$. For the dimer model on an infinitely
long cylinder of odd circumference $\mathcal{N}$ one can obtained
from Eqs.(\ref{freeenergystrip}), (\ref{Aperiod}) and
(\ref{2N1per}) that the central charge $c$ and the highest
conformal weight $\Delta$ can take values $c = 1$ and
$\Delta=1/16$. It turns out in this case that another consistent
conformal description exists, with $c=-2$ and $\Delta=0$
\cite{izmailian2005}. In particular, it has been shown (for more
details see \cite{izmailian2005}) that although the dimer model is
originally defined on a cylinder with odd circumference
$\mathcal{N}$, it shows the finite-size corrections expected on a
strip and must really be viewed as a model on a strip.

Thus from the finite size analyzes we can see that two conformal
field theories with the central charges $c=1$ and $c=-2$ can be used
to  described the dimer model on the triangular lattice. But since
the general triangular lattice model is critical in the square
lattice limit and the dimer model on the square lattice belongs to
$c=-2$ universality class \cite{izmailian2005}, we come to the
conclusion that the dimer model on the triangular lattice can also be
described by conformal field theory having central charge $c=-2$.

%%%%%%%%%%%%%%%%%%%%%%%%%%%%%%%%%%%%%%%%%%%%%%%%%%%%%%%%%%%%%%%%%%%%%%%%%%%%%%%%%%%%%%%%%%%%%%%%%%%%%%%%%%%%%%%%%%%%%%%%%%%%%%%%%%%%%
\section{Specific heat near the critical point}
\label{sec_heat}
%%%%%%%%%%%%%%%%%%%%%%%%%%%%%%%%%%%%%%%%%%%%%%%%%%%%%%%%%%%%%%%%%%%%%%%%%%%%%%%%%%%%%%%%%%%%%%%%%%%%%%%%%%%%%%%%%%%%%%%%%%%%%%%%%%%%%

Let us now consider the behavior of the specific heat near the
critical point. The specific heat $C(t,\mathcal{M},\mathcal{N})$
of the dimer model on $\mathcal{M} \times \mathcal{N}$ triangular
lattice is defined as
\begin{eqnarray}
C(t,\mathcal{M},\mathcal{N})&=&-\frac{\partial^2 }{\partial
t^2}f(t,\mathcal{M},\mathcal{N}), \label{specificheat}
\end{eqnarray}
where $f(t,\mathcal{M},\mathcal{N})$ is free energy of the system
\begin{eqnarray}
f(t,\mathcal{M},\mathcal{N})&=&-\frac{1}{\mathcal{S}}\ln
Z_{\mathcal{M},\mathcal{N}}(t), \label{freeenergyMN}
\end{eqnarray}
and where $\mathcal{S}=\mathcal{M} \mathcal{N}$ is the lattice area.

The pseudocritical point $t_{\rm{pseudo}}$ is the value of the
temperature at which the specific heat has its maximum for finite
$\mathcal{M} \times \mathcal{N}$ lattice. One can determine this
quantity as the point where the derivative of
$C(t,\mathcal{M},\mathcal{N})$ vanishes. The pseudocritical point
approaches the critical point $t_c=0$ as $L \to \infty$ in a
manner dictated by the shift exponent $\lambda$,
\begin{eqnarray}
|t_{\rm{pseudo}}-t_c| \sim L^{-\lambda}. \label{lamb}
\end{eqnarray}
where $L = \sqrt{\mathcal{S}}$ is the characteristic size of the
system. The coincidence of $\lambda$ with $1/\nu$, where $\nu$ is
the correlation lengths exponent, is common to most models, but it
is not a direct consequence of finite-size scaling and is not
always true.

Since the expression for the partition function is different for
even $\mathcal{N}$ and odd $\mathcal{N}$, we  consider these two
cases separately.

Let us start with case of odd $\mathcal{N}$ ($\mathcal{N}=2N+1$).
Expanding the expression (\ref{specificheat}) about the critical
point $t=0$ with the help of Eqs.(\ref{ident1}), (\ref{ident2})
and (\ref{eo}) yields
\begin{eqnarray}
C(t,\mathcal{M},\mathcal{N}) = C(0,\mathcal{M},\mathcal{N}) + t\;
C^{(1)}(0,\mathcal{M},\mathcal{N})+\frac{t^2}{2}C^{(2)}(0,\mathcal{M},\mathcal{N})+O(t^3),
\label{expansion specific heat1}
\end{eqnarray}
where $C(0,\mathcal{M},\mathcal{N})$ is the critical specific
heat, and ${\displaystyle{C^{(n)}(0,\mathcal{M},\mathcal{N})
\equiv \frac{d^n}{dt^n}C(t,\mathcal{M},\mathcal{N})|_{t=0}}}$. We
have
\begin{eqnarray}
\mathcal{S}\;C(0,\mathcal{M},\mathcal{N}) &=&
\frac{Z_{0,1/2}^{(2)}(0,\mathcal{M}/2,\mathcal{N})}{Z_{0,1/2}(0,\mathcal{M}/2,\mathcal{N})}-\left(\frac{Z_{0,0}^{(1)}(0,\mathcal{M}/2,\mathcal{N})}{Z_{0,1/2}
(0,\mathcal{M}/2,\mathcal{N})}\right)^2,
 \label{C0}\\
\mathcal{S}\;C^{(1)}(0,\mathcal{M},\mathcal{N})&=&
\frac{Z_{0,0}^{(3)}(0,\mathcal{M}/2,\mathcal{N})}{Z_{0,1/2}(0,\mathcal{M}/2,\mathcal{N})}
-
3\frac{Z_{0,0}^{(1)}(0,\mathcal{M}/2,\mathcal{N})Z_{0,1/2}^{(2)}(0,\mathcal{M}/2,\mathcal{N})}
{Z_{0,1/2}^2(0,\mathcal{M}/2,\mathcal{N})}
\nonumber \\
 &+&2\left(\frac{Z_{0,0}^{(1)}(0,\mathcal{M}/2,\mathcal{N})}{Z_{0,1/2}(0,\mathcal{M}/2,\mathcal{N})}\right)^3,\label{C1}\\
\mathcal{S}\;C^{(2)}(0,\mathcal{M},\mathcal{N})&=&\frac{Z_{0,1/2}^{(4)}(0,\mathcal{M}/2,\mathcal{N})}{Z_{0,1/2}(0,\mathcal{M}/2,\mathcal{N})}
+12
\left(\frac{Z_{0,0}^{(1)}(0,\mathcal{M}/2,\mathcal{N})}{Z_{0,1/2}(0,\mathcal{M}/2,\mathcal{N})}\right)^2\frac{
Z_{0,1/2}^{(2)}(0,\mathcal{M}/2,\mathcal{N})}{Z_{0,1/2}(0,\mathcal{M}/2,\mathcal{N})}
\nonumber\\
&-&3\left(\frac{Z_{0,1/2}^{(2)}(0,\mathcal{M}/2,\mathcal{N})}{Z_{0,1/2}(0,\mathcal{M}/2,\mathcal{N})}\right)^2-
4\frac{Z_{0,0}^{(1)}(0,\mathcal{M}/2,\mathcal{N})Z_{0,0}^{(3)}(0,\mathcal{M}/2,\mathcal{N})}{Z_{0,1/2}^2(0,\mathcal{M}/2,\mathcal{N})}\nonumber\\
&-&6\left(\frac{Z_{0,0}^{(1)}(0,\mathcal{M}/2,\mathcal{N})}{Z_{0,1/2}(0,\mathcal{M}/2,\mathcal{N})}\right)^4.\label{C2}
\end{eqnarray}
From Eq.(\ref{expansion specific heat1}), the first derivative of
the specific heat on a finite lattice near the infinite volume
critical point can be found, and it seen to vanish when
\begin{eqnarray}
t_{\rm{pseudo}} =
-\frac{C^{(1)}(0,\mathcal{M},\mathcal{N})}{C^{(2)}(0,\mathcal{M},\mathcal{N})}.
\label{expansionmu}
\end{eqnarray}
To find the exponent $\lambda$ one need the finite-size
corrections to $C^{(1)}(0,\mathcal{M},\mathcal{N})$ and
$C^{(2)}(0,\mathcal{M},\mathcal{N})$. The exact asymptotic
expansions of $C^{(1)}(0,\mathcal{M},\mathcal{N})$ and
$C^{(2)}(0,\mathcal{M},\mathcal{N})$ can be found along the same
lines as in Ref.\cite{izmailian2007} and leading finite size
behavior is
\begin{eqnarray}
C^{(1)}(0,\mathcal{M},\mathcal{N})& \sim & \sqrt{\mathcal{S}},
\label{expansiont1}\\
C^{(2)}(0,\mathcal{M},\mathcal{N})& \sim & \mathcal{S}.
\label{expansiont2}
\end{eqnarray}
Expressions Eq.(\ref{expansiont1}) and (\ref{expansiont2}) now
gives the FSS of the pseudocritical point to be
\begin{eqnarray}
t_{\rm{pseudo}} \sim L^{-1}, \label{expansionmupseudo}
\end{eqnarray}
where $L = \sqrt{\mathcal{S}}$ is the characteristic size of the
system. Thus from Eqs. (\ref{lamb}) and (\ref{expansionmupseudo})
we find that the shift exponent is $\lambda=1$ for odd
$\mathcal{N}$. In Fig. 2(a) and 2(b) we plot the $t$ dependence of
the specific heat for $16 \times 17$ lattice. We can see from Fig.
2(b) that the position of the specific heat peak is shifted from
zero.

For even $\mathcal{N}$ ($\mathcal{N}=2N$), one can see from
Eqs.(\ref{ee}), (\ref{ident1}), (\ref{ident2}),
(\ref{specificheat}) and (\ref{freeenergyMN}) that the partition
function $Z_{{\cal M},{\cal N}}(t)$ and the specific heat
$C(t,\mathcal{M},\mathcal{N})$ are an even function with respect
to its argument $t$
\begin{eqnarray}
C(t,\mathcal{M},\mathcal{N}) = C(0,\mathcal{M},\mathcal{N})
+\frac{t^2}{2}C^{(2)}(0,\mathcal{M},\mathcal{N})+\frac{t^4}{4!}C^{(4)}(0,\mathcal{M},\mathcal{N})+O(t^6).
\label{expansion specific heat}
\end{eqnarray}
The partition function given by Eq. (\ref{PartitionFunctionDimer})
is a polynomial function of its argument $t$. In the case of even
${\cal M}$ and ${\cal N}$ the partition function is an even
function with respect of its argument $t$ and hence the number of
diagonal bonds $n_t$ also should be even. There is also a simple
geometrical explanation why, in the case of even-even lattices,
the number of diagonal dimers should be even. Let us consider an
even-even ($2{\cal M} \times 2{\cal N}$) lattice with periodic
boundary conditions in the vertical and horizontal directions.
Such a lattice can be divided into two sublattice A and B, as
shown in Fig.~3. Each sublattice consists of $2{\cal M}{\cal N}$
sites in a such way that every horizontal or vertical edge
connects a site in sublattice A to one in sublattice B, while
diagonal edges connect two sites within sublattice A or two sites
within sublattice B. Note that such a division into A and B
sublattices is impossible for even-odd ($2{\cal M} \times 2{\cal
N}+1$) lattices with periodic boundary conditions, since in that
case, one can always find a horizontal or vertical edge which
connects two sites in the same sublattice A or B. Thus, in the
even-even case, each horizontal or vertical dimer occupies one
site from sublattice A and another site from sublattice B, while
diagonal dimer occupies two sites from sublattice A or B. If one
diagonal dimer occupies two sites from sublattice A one should
have another diagonal dimer which occupies two sites of sublattice
B in order to insure that remaining sites can be occupied by
horizontal and vertical dimers. Thus, in the case of even M and N,
only even number of diagonal bonds are allowed. Similar
geometrical considerations for the dimer model on $({\cal M}
\times {\cal N})$ lattice with free boundary conditions lead to
the conclusion that for both even-even and even-odd lattices the
number of diagonal bonds should be even.

Thus the first derivative of $C(t,\mathcal{M},\mathcal{N})$
vanishes exactly at
\begin{eqnarray}
t_{\rm{pseudo}}=0. \label{expansionmu0}
\end{eqnarray}
In Fig. 4(a) and 4(b) we plot the $t$ dependence of the specific
heat for a $16 \times 16$ lattice. We can see from Fig. 3(b) that
the position of the specific heat peak $t_{pseudo}$ is equal
exactly to zero.
\begin{figure}
\epsfxsize=70mm \vbox to2.0in{\rule{0pt}{2.0in}}
\includegraphics{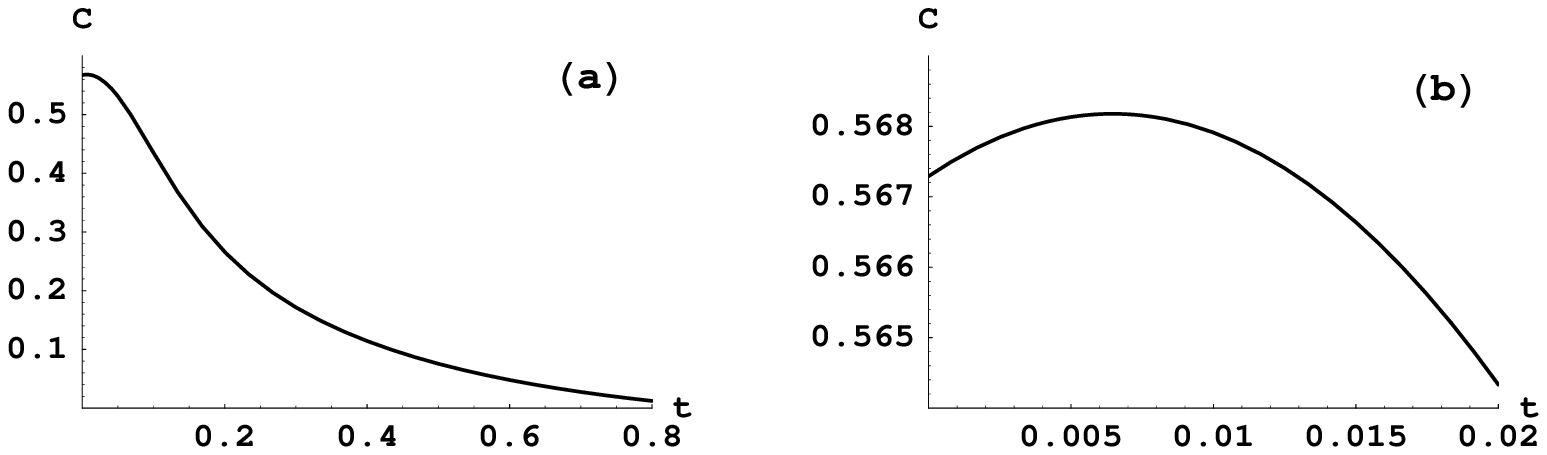}\caption{(a) The behavior of the specific heat
on ${\cal M} \times {\cal N}$ lattices with even ${\cal M}=16$ and
odd ${\cal N}=17$. (b) The behavior of the specific heat on the
same lattice for small $t$.} \label{fig2}
\end{figure}
\begin{figure}
\epsfxsize=70mm \vbox to3.0in{\rule{0pt}{3.0in}}
\includegraphics{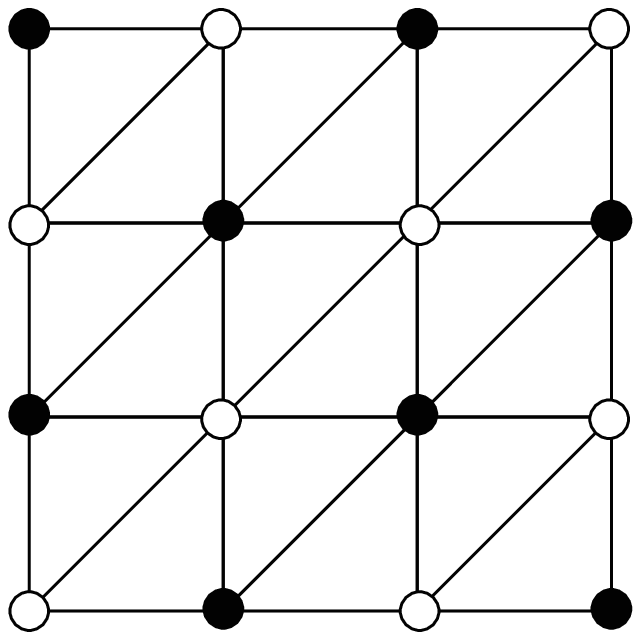}\caption{Division of lattice on two sublattice A
(black circles) and B (white circles).} \label{fig3}
\end{figure}
%The inset in a zoom in the region of the maximum.
\begin{figure}
\epsfxsize=70mm \vbox to2.0in{\rule{0pt}{2.0in}}
\includegraphics{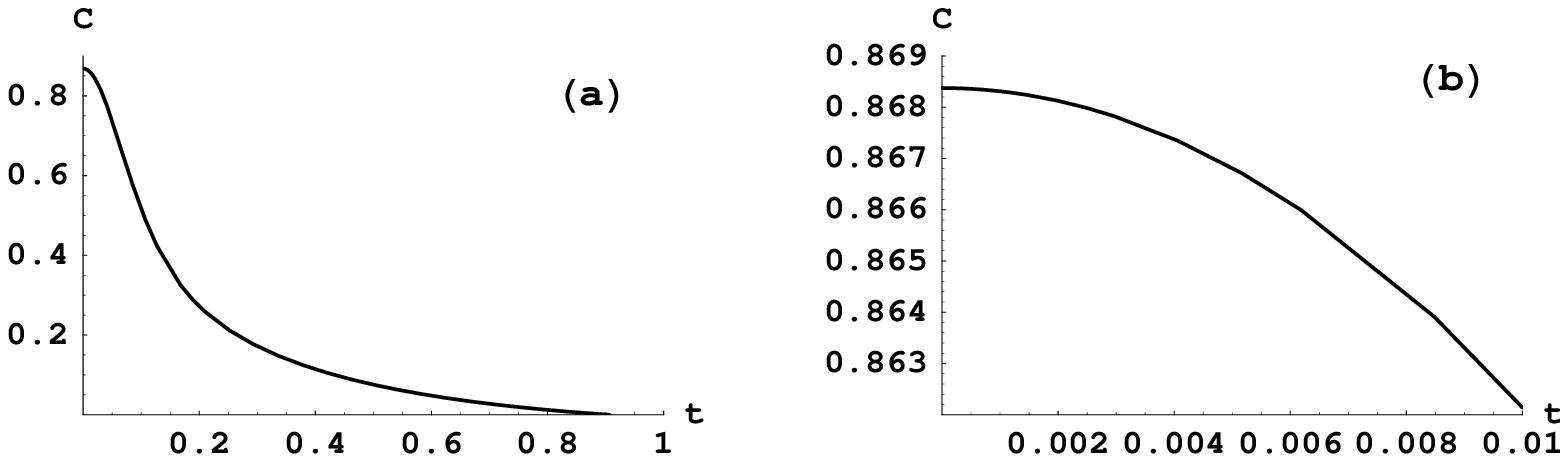}\caption{(a) The behavior of the specific heat
on $M \times N$ lattices with even $M=16$ and $N=16$. (b) The
behavior of the specific heat on the same lattice for small $t$.}
\label{fig4}
\end{figure}
Therefore the maximum of the specific heat (the pseudocritical
point $t_{\rm{pseudo}}$)  always occurs at vanishing reduced
temperature for any finite  $M \times 2N$ lattice and coincides
with the critical point $t_c$ at the thermodynamic limit. From
Eqs. (\ref{lamb}) and (\ref{expansionmu0}) we find that the shift
exponent is $\lambda=\infty$ for even $\mathcal{N}$.

Thus we have found that the shift exponent $\lambda$ for the
specific heat depend in a crucial way on the parity of the number
of lattice sites along the lattice axis $\mathcal{N}$. For odd
$\mathcal{N}$ we obtain for the shift exponent $\lambda=1$, while
for even $\mathcal{N}$ we have found that the shift exponent
$\lambda$ is infinity ( $\lambda=\infty$).

\section{Conclusion}
\label{conclusion} We analyze the partition function of the dimer
model on $\mathcal{M} \times \mathcal{N}$ triangular lattice
wrapped on torus obtained by Fendley, Moessner and Sondhi
\cite{Moesner}. From a finite-size analysis we have found that the
dimer model on the triangular lattice can be described by
conformal field theory having central charge $c=-2$. Thus we have
shown that the dimer model on the triangular lattice belongs to
the same universality class as the dimer model on the square
lattice, while the dimer model on the honeycomb lattice belongs to
another $c=1$ universality class. In addition, we have found that
the shift exponent $\lambda$ depends in a crucial way on the
parity of the number of lattice sites along the lattice axis
$\mathcal{N}$: for odd $\mathcal{N}$ we obtain
 $\lambda=1$, while for even $\mathcal{N}$ we have
found that $\lambda=\infty$. In the former case, therefore, the
finite-size specific-heat pseudocritical point is size dependent,
while in the latter case it coincides with the critical point of
the thermodynamic limit. This adds to the catalog of anomalous
circumstances where the shift exponent is not coincident with the
correlation-length critical exponent. The present circumstance
manifests the additional feature that the shift-exponent is
boundary-condition dependent.

\section{Acknowledgment}

One of us (N.Sh.I) thanks the Statistical Physics Group at the
Departamento de Ciencias Exatas, Universidade Federal de Lavras,
Brazil, for hospitality during completion of this work. This work
was supported by the EU Programme FP7-People-2010-IRSES (Project
No 269139) and partially supported by FAPEMIG (BPV-00061-10).

%%%%%%%%%%%%%%%%%%%%%%%%%%%%%%%%%%%%%%%%%%%%%%%%%%%%%%%%%%%%%%%%%%%%%%%%%%%%%%%%%%%%%%%%%%%%%%%%%%%%%%%%%%%%%%%%%%%%%%%%%%%%%%%%%%%%%

\end{document}